\documentclass[journal,twoside]{IEEEtran}
\ifCLASSINFOpdf
\else
\fi

\usepackage{hyperref}
\usepackage{amsfonts, amsmath, amssymb}
\usepackage{graphicx}
\usepackage{afterpage}
\usepackage[dvipsnames]{xcolor}
\usepackage{multirow}
\usepackage{cite}

\newcommand{\norm}[1]{\lVert#1\rVert}

\begin{document}

%
\title{Graph convolutional networks enable fast hemorrhagic stroke monitoring with electrical impedance tomography}
%
%
%
\author{J.~Toivanen, V.~Kolehmainen, A.~Paldanius, A.~H\"anninen, A. Hauptmann and S.~J. Hamilton

\thanks{SH was supported by the National Institute Of Biomedical Imaging and Bioengineering of the National Institutes of Health under Award Number R21EB028064. The content is solely the responsibility of the authors and does not necessarily represent the official views of the National Institutes of Health.  JT, VK, AP, AH, and AH were supported in part by the Research council of Finland (Projects No. 359433, 353084, 353093, Finnish Centre of Excellence in Inverse Modeling and Imaging; and Projects No. 358944, 359186, Flagship of Advanced Mathematics for Sensing Imaging and Modelling; Project No. 338408, Academy Research Fellow), the Jane and Aatos Erkko Foundation and Neurocenter Finland.}
\thanks{J. Toivanen, V. Kolehmainen, and A. Hänninen are with the Department of Technical Physics, University of Eastern Finland, FI-70210 Kuopio, Finland.}%
\thanks{A. Paldanius is with the Faculty of Medicine and Health Technology, Tampere University, Tampere, Finland.}%
\thanks{S.J. Hamilton are with the Department of Mathematical and Statistical Sciences; Marquette University, Milwaukee, WI 53233 USA, (e-mail:\texttt{(sarah.hamilton@mu.edu)}).}%
\thanks{A. Hauptmann is with the Research Unit of Mathematical Sciences, University of Oulu, Oulu, Finland and with the Department of Computer Science, University College London, London, United Kingdom.}%

\thanks{Manuscript submitted: October 24, 2024.}} 

%
%

\markboth{Journal of \LaTeX\ Class Files}
{XXX et al. \MakeLowercase{\textit{et al.}}: Bare Demo of IEEEtran.cls for IEEE Journals}
%



\maketitle

\begin{abstract}
~{\it Objective:} To develop a fast image reconstruction method for stroke monitoring with electrical impedance tomography with image quality comparable to computationally expensive nonlinear model-based methods. {\it Methods:} A post-processing approach with graph convolutional networks is employed.  Utilizing the flexibility of the graph setting, a graph U-net is trained on linear difference reconstructions from 2D simulated stroke data and applied to fully 3D images from realistic simulated and experimental data.  An additional network, trained on 3D vs. 2D images, is also considered for comparison.  {\it Results:} Post-processing the linear difference reconstructions through the graph U-net significantly improved the image quality, resulting in images comparable to, or better than, the time-intensive nonlinear reconstruction method (a few minutes vs. several hours).  {\it Conclusion:} Pairing a fast reconstruction method, such as linear difference imaging, with post-processing through a graph U-net provided significant improvements, at a negligible computational cost.  Training in the graph framework vs classic pixel-based setting (CNN) allowed the ability to train on 2D cross-sectional images and process 3D volumes providing a nearly 50x savings in data simulation costs with no noticeable loss in quality.  {\it Significance:} The proposed approach of post-processing a linear difference reconstruction with the graph U-net could be a feasible approach for on-line monitoring of hemorrhagic stroke.



\end{abstract}



%
\IEEEpeerreviewmaketitle

\section{Introduction}\label{sec:Intro}

\begin{figure}[t!]
\linethickness{.3mm}
    \centering
    \includegraphics[width=\columnwidth]{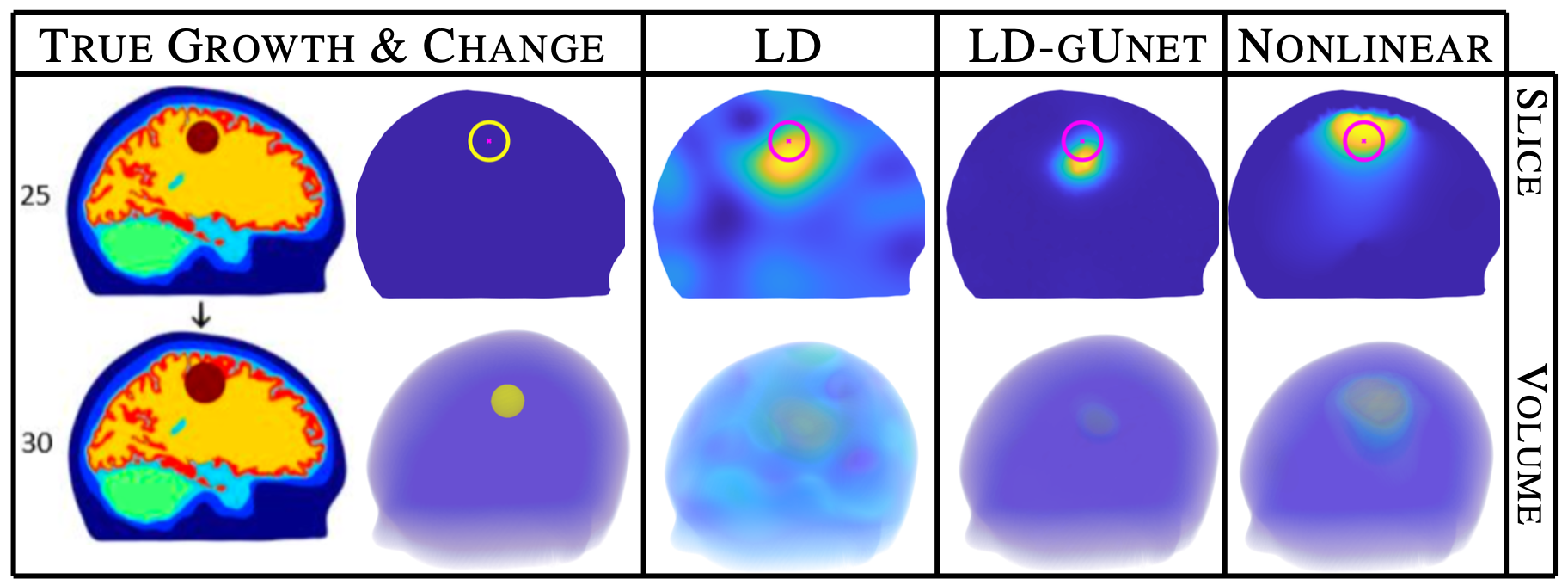}
\caption{Demonstration of the proposed method.  Fast, blurry linear difference (LD) reconstructions are post-processed with a graph U-net, at negligible cost, resulting in images of comparable quality to computationally expensive nonlinear methods.}
    \label{fig:hammer}
\end{figure}

\IEEEPARstart{T}{he} focus of this work is the use of electrical impedance tomography (EIT) for the challenging task of stroke monitoring.  Here we propose the use of graph U-nets to drastically improve the quality of fast, linearized, stroke monitoring imaging, at negligible cost, to result in improved images of quality comparable to computationally expensive reconstruction methods (see Figure~\ref{fig:hammer}).  The learning is performed on the underlying graph structure of the image to allow for training on 2D cross-sectional brain images and subsequent processing of 3D volumes, exploiting the dimension-free nature of graph networks in contrast to classic CNN U-nets.  This results in a 50x computational savings in data simulation alone.

Worldwide, strokes are a leading cause of death and disability 
\cite{katan2018global}. Approximately 70\% of strokes are ischemic, resulting from a clot or blockage of blood-flow in the brain, and 30\% hemorrhagic (bleeds). The most common type of hemorrhagic stroke is an intracerebral hemorrhage (ICH), which is usually caused by disruption of cerebral arteries leading to bleeding into the brain tissues.  In the early stage of ICH, within 1-2 days from the onset, possible expansion of the hematoma and re-bleeding form a formidable risk to the patient \cite{greenberg20222}. Additionally, patients with thrombolysis treated acute ischemic stroke are prone to a
secondary cerebral hemorrhage \cite{sussman2013hemorrhagic}. Such secondary hemorrhage in ischemic stroke patients as well as hematoma expansion or re-bleeding in ICH patients should be detected immediately. 


The follow-up of ICH in the intensive care facility is mainly based on monitoring of clinical signs and symptoms \cite{greenberg20222}. However, vigilant observation of subtle neurological signs of the patient can be challenging, particularly in intubated and sedated patients. Currently the most reliable way to monitor ICH is through repeated computerized tomography (CT) scanning \cite{greenberg20222}. However, deciding the timing of the control CT imaging is itself already a difficult problem, and CT provides only snapshots of the bleeding at the times of the control imaging. Further, transporting a patient from intensive care to radiology and back repeatedly is demanding and can be dangerous for the patient. Therefore, there is need for an on-line patient monitoring system that could be used at the bedside for detecting possible changes in the intracerebral hemorrhage. 

One promising bedside monitoring technology is electrical impedance tomography (EIT). In EIT small alternating currents are injected through electrodes attached to the patient’s scalp and the resulting voltages are measured on the electrodes. The electrical current and voltage data is then used to reconstruct a three-dimensional image of the electrical conductivity of the brain, with hemorrhage occuring as higher conductivity compared to normal brain tissue due to the high conductivity of blood. This type of absolute EIT imaging has been studied for early differentiation of stroke types, see e.g. \, \cite{malone2014, goren2018, horesh2006thesis, candiani2021,agnelli2020classification, mcdermott2020, candiani2022}. However, stroke differentiation using EIT is very challenging as one should not overlook even a tiny amount of bleeding in the clinical decision making, and further technical challenges arise from the sensitivity of the absolute EIT problem to modeling errors, such as inaccurately known electrode locations and head shape.  In a continuous bedside monitoring scenario, the EIT measurement is taken repeatedly at regular time intervals and the conductivity changes of the brain between the consecutive EIT measurements is estimated.
This type of difference imaging approach is promising for on-line stroke monitoring and detection of secondary hemorrhages as the measurements have good sensitivity for changes in the amount of highly conductive blood, as suggested by simulation \cite{shi2021sparse}, animal \cite{xu2010} and human studies \cite{dai2013, yang2019}, and the difference imaging setup is also significantly less prone to the modeling errors.

In the envisioned stroke monitoring scenario, EIT measurements of the patient with a confirmed ICH is repeated at regular intervals, e.g., every 2-10 minutes, and an image reconstruction algorithm is used to compute 3D images of the conductivity change in the brain between the measurements. Detected changes indicative of changes in the ICH would indicate an immediate need for control CT imaging or medical intervention. The EIT based stroke imaging could also be used to monitor for secondary hemorrhages in the treatment of ischemic stroke patients. Stroke monitoring can also utilize the patient's CT image, routinely taken at admission, to generate a patient-specific 3D computational mesh and improved regularization model \cite{toivanen2024simulation}.

Classic image reconstruction algorithms, such as linearization-based approaches, for stroke monitoring suffer from artifacts, sub-optimal resolution and localization of inclusions. Improved detection of hemorrhage growth can be obtained by non-linear optimization-based approaches such as the nonlinear monitoring algorithm presented in \cite{toivanen2024simulation,Toivanen21}.  However, these approaches lead to long computation times in 3D EIT, impractical for on-line monitoring in intensive care.

In recent years, machine learning approaches have been utilized to overcome the above computational challenges in EIT \cite{tanyu2023electrical}. There are three distinct approaches: direct learned reconstructions; learned post-processing; and iterative model-based reconstructions. In the first approach, a network is trained to directly form an image from the measured data \cite{li2017image,chen2022mmv}. This approach, while fast, often does not generalize well to testing data that is different from the training data or strongly corrupted by noise. In the second case, a neural network is trained to improve an initial reconstruction, such as a D-bar reconstruction or an early iterate of a variational method, with a post-processing network \cite{Hamilton2018_DeepDbar,wei2019dominant,hamilton2019beltrami,chen2021structure}. In the third case, a neural network is trained to learn more efficient update steps in an iterative algorithm to speed-up the reconstruction \cite{herzberg2021graph}. The two latter approaches have been shown to improve robustness to noise and out-of-distribution test data. Additionally, the explicit modeling allows for more flexibility with respect to measurement geometries. Nevertheless, a common limitation exists with respect to accessibility of training data, necessary for supervised training.
This is especially a problem in 3D where very little measurement data is available, and even accurate simulations to create reference data can take a considerable amount of time. In this work we address the problem of training data creation in 3D by utilizing an efficient training in 2D with the ability to generalize to 3D. This can be achieved by using a training geometry and dimension-independent graph convolutional network.

The generalization capability of graph neural networks in the context of EIT has been recently presented in \cite{herzberg2023domain} and \cite{Dong2024Graph}. In the former, a graph based U-net was introduced that was trained on early iteration total variation (TV) reconstructions. In a proof-of-concept experiment the authors showed that the network can be applied to 3D input data coming from two very different reconstruction methods, Levenberg-Marquardt and the $t^{\mbox{\tiny \textbf{exp}}}$ method (a direct method based on complex geometrical optics solutions).  
However, how well this approach of training on data coming from a lower dimension (e.g. 2D) and processing higher dimensional data (e.g. 3D) through a trained graph neural network works remained unexplored. Consequently, in this work, we present, for the first time, a detailed study of the effect of training in 2D and processing on 3D data, for the example of stroke monitoring with EIT.  We demonstrate that, for our problem, the results of training on 2D data are on par, if not superior to, training on the 3D data itself.  As the computational cost of generating complex 3D training data is high, and the subsequent neural network training longer, such time savings could be advantageous in complex settings.  We focus here on graph U-nets and train them to post process linearized difference imaging reconstructions, using both 2D and 3D simulated training data. The approaches are tested with out-of-distribution 3D stroke monitoring data coming from a detailed six-layer anatomical head model~\cite{Paldanius21}, and experimental data from a 3D-printed head model. The results demonstrate that the graph U-net post-processing significantly improves the detection of hemorrhage growth for the linearized algorithm, resulting in performance similar to the non-linear monitoring algorithm \cite{toivanen2024simulation} but
with a short, and clinically feasible, computation time. 

The remainder of the paper is organized as follows.  Section~\ref{sec:Methods} describes the learning problem: a brief overview of electrical impedance tomography, the linear difference (LD) reconstruction method to be used as input images in the training, and the nonlinear (MO) reconstruction algorithm used as a quality standard, and the graph U-net framework.  Details of the simulation of training data, and out-of distribution testing data, are provided in Section~\ref{sec:sampelsMetrics} along with a description of the metrics that will be used to assess the quality across the reconstruction methods.  Results for out-of-distribution testing data on a complex head model, as well as experimental data, are presented in Section~\ref{sec:results}.  Further comparison of the networks, and discussion of the results, are described in Section~\ref{sec:discussion} and conclusions drawn in Section~\ref{sec:conclusion}.

\section{Methods}\label{sec:Methods}

\subsection{Electrical impedance tomography}\label{sec:eit}
This section gives an overview of the forward model of EIT and the setup for the stroke monitoring application. We briefly review  the linearized reconstruction approach, which we post process with the graph U-net, and the non-linear difference imaging algorithm \cite{Toivanen21,toivanen2024simulation} which we use as reference for assessing the quality across images.
\subsubsection{Forward model}\label{sec:forwardmodel}

As we consider both 2D and 3D reconstructions, we consider the domain $\Omega \subset \mathbb{R}^d \, (d=2,3)$ and model the $L$ electrodes attached to the boundary as subsets (segments/patches) of the exterior boundary,  denoted $e_\ell \subset \partial \Omega$, $l = 1, 2, \ldots, L$.  In the EIT measurement, $P$ current patterns $I^{(j)} \in \mathbb{R}^L$, $j = 1, 2, \ldots , P$, are consecutively applied through the electrodes, and the corresponding voltages $U^{(j)} \in \mathbb{R}^L$ are measured on all electrodes. Here $I_\ell^{(j)}$ and $U_\ell^{(j)}$ denote the applied current and measured voltage from the $j^{th}$ current pattern on the $\ell^{th}$ electrode for $\ell = 1, 2, \ldots, L$. Based on the conservation of charge, and our choice of electric potential ground, we have
\begin{equation}
 \sum_{\ell=1}^L {I_\ell^{(j)}} = 0 \qquad \qquad \sum_{\ell=1}^L {U_\ell^{(j)}} = 0.
 \label{equ:ForwardModel1} 
\end{equation}
The voltages $U_\ell^{(j)}$ are boundary measurements of the interior electromagnetic potential $u^{(j)}(x)$ that is modeled with the conductivity equation
\begin{align}
\nabla \cdot (\sigma(x) \nabla u^{(j)}(x) )= 0,&\  x \in \Omega,
\label{equ:ForwardModel2}
\end{align}
where $\sigma (x)$ is the the conductivity and the boundary conditions of the complete electrode model (CEM) \cite{Cheng1989,Somersalo1992} for $j = 1, \dots, P$ and $\ell=1, \dots, L$, are given by 
\begin{align}
u^{(j)}(x)+z_{\ell} \sigma(x) \frac{\partial u^{(j)}(x)}{\partial n} &= U^{(j)}_{\ell}, \ \ x \in e_{\ell}
\label{equ:ForwardModel3} \\  
\int_{e_{\ell}}\sigma(x) \frac{\partial u^{(j)}(x)}{\partial n} \,\mathrm{d}S&=I^{(j)}_{\ell},
\label{equ:ForwardModel4}  \\ 
\sigma(x) \frac{\partial u^{(j)}(x)}{\partial n}&=0, \ \  x\in
\partial\Omega \backslash \bigcup_{\ell=1}^{L} e_{\ell}.
\label{equ:ForwardModel5} 
\end{align}
Here $z_\ell$ denotes the contact impedance between the electrode $e_{\ell}$ and the body $\Omega$, and $n$ the outward unit normal vector on the boundary $\partial\Omega$.  Proof for the existence and uniqueness of the solution 
for the variational formulation of the model \eqref{equ:ForwardModel1}-\eqref{equ:ForwardModel5} can be found in \cite{Somersalo1992}.

In this paper, the numerical solution of the model  \eqref{equ:ForwardModel1}-\eqref{equ:ForwardModel5}  is based on the finite element method (FEM); for details of the implementation see \cite{vauhkonen98,Kaipio2000stat}. In the following, we denote the FEM based solution for a single current pattern $I^{(j)}\in\mathbb{R}^L$ by $\sigma \mapsto U(\sigma;I^{(j)}) \in \mathbb{R}^L$ where we consider a discretized version of $\sigma$, such that $\sigma = \sum_{h=1}^N \sigma_h \varphi_h$, where $\sigma_h \in \mathbb{R}_+$ are the nodal coefficients
and $\varphi_h \in H^1(\Omega)$, $h=1, \dots, N$, are the piecewise linear basis functions of a FE mesh of $\Omega$. 

The measurement noise $e \in \mathbb{R}^{P\cdot L}$ is modeled as additive noise, leading to the measurement model
\begin{equation}
  \label{equ:MeasurementModel} 
  V = U(\sigma) + e,
\end{equation}
where  $V\in\mathbb{R}^{P\cdot L}$ is the vector of the measured noisy voltages for all applied current patterns and $U(\sigma) = (U(\sigma;I^{(1)}), \dots, U(\sigma;I^{(P)}))^{\rm T}$ $\in \mathbb{R}^{P\cdot L}$.

\subsubsection{Monitoring of hemorrhagic stroke using EIT}

The growth of an ICH, or occurrence of a secondary hemorrhage, is expected to present as a localized change in the brain conductivity due to an increased volume of highly conducting blood at the hemorrhage location. Thus, the
aim in EIT monitoring is to monitor the status of the hemorrhage by reconstructing changes of the
conductivity,
\begin{equation}
    \label{equ:ds}
    \delta \sigma = \sigma_2 - \sigma_1,
\end{equation}
based on EIT measurements $V_1$ and $V_2$ taken at times $t_1$ and $t_2$, roughly 2-10 minutes apart. 
The acquisition of a single EIT measurement is very fast, typically multiple frames per second. 
The acquisition of the single frame is short compared to the changes in the hemorrhage status and the measurement is also fast compared to the time between consecutive measurements, implying that it is
feasible to model the measurements separately with the stationary 
model
\begin{align}
    \label{equ:V1And2}
    V_k = U(\sigma_k) + e_k ,\quad k = 1, 2.
\end{align}

\subsubsection{Linear difference imaging algorithm}\label{sec:methods-linear}
The linear difference (LD) imaging algorithm, see e.g. \cite{barber1987fast, Bagshaw2003},  is based on a linearization of the forward model \eqref{equ:V1And2} and aims to reconstruct the change in conductivity $\delta \sigma$ between $V_1$ and $V_2$ 
by solving the regularized linear least-squares problem
\begin{equation}
\label{equ:minLD}
  \hat{\delta\sigma} = \arg \min_{\delta \sigma} \left\{ \norm{L_{\delta e}(\delta V - J\delta\sigma) }^2 + \norm{ R\delta\sigma}^2 \right\},
\end{equation}
where $\delta V = V_2 - V_1$, $\delta e = e_2 - e_1$, $L_{\delta e}$ is the Cholesky factor of the noise precision matrix of $\delta e$ so that $L_{\delta e}^{\rm T}L_{\delta e} = \Gamma_{\delta e}^{-1} = (\Gamma_{e_1} + \Gamma_{e_2})^{-1}$, the Jacobian matrix $J$ of $U(\sigma)$ is evaluated at a 
linearization point $\sigma_0$, and $R$ is a regularization matrix.

In this paper, as in \cite{Toivanen21, toivanen2024simulation}, the linearization point $\sigma_0$ was obtained by solving a non-linear least squares fitting problem for the best fitting spatially constant conductivity using the measurement data $V_1$. 
The regularization matrix $R$ was constructed utilizing a distance-based correlation model \cite{lieberman2010parameter} resulting in regularization that promotes spatially smooth conductivity changes. For implementation details, see \cite{Toivanen21, toivanen2024simulation}.  Figure~\ref{fig:hammer} (column 2) shows a sample reconstruction produced using this LD method 
in 49 seconds.

\subsubsection{Nonlinear stroke monitoring algorithm}

The nonlinear stroke monitoring (MO) algorithm
\cite{toivanen2024simulation} is used as reference for evaluating the image quality of the learning-based post processing of LD images. The MO method utilizes non-linear difference imaging \cite{Liu2015ipi}, parallel level sets regularization \cite{kolehmainen2019ipi} and a priorconditioned least squares algorithm \cite{Arridge14LSQR, Harhanen15LSQR}. The algorithm utilizes prior information from the patient CT, which is taken for diagnosis of the stroke at the time of patient admission to the hospital, for the construction of the patient specific 3D mesh and parallel level set regularization, favoring similar alignment of level-set lines in the CT and EIT reconstruction of the baseline conductivity $\sigma_1$ at time $t_1$.

The algorithm also allows utilization of a region of interest (ROI) constraint for the conductivity change
\begin{equation}
    \label{equ:ROIConst}
    \textrm{supp}(\delta \sigma) = \Omega_{\textrm{ROI}} \subseteq \Omega.
\end{equation}
The ROI can be chosen based on the patient CT image and, for ICH monitoring, a natural catch-all choice is to use the full brain volume. Given the selected ROI, the conductivity at the later measurement time $t_2$ is modeled as
\begin{equation}
    \label{equ:s2}
    \sigma_2 = \sigma_1 + K \delta \sigma,
\end{equation}
where $K$ is an extension mapping that zero-extends the conductivity change from the ROI to the whole domain. In case no ROI is used, $K$ equals the identity matrix with the same dimension as $\sigma_1$. 

The change in conductivity $\delta \sigma$ between the two measurements $V_1$ and $V_2$ is estimated by the non-linear optimization problem
\begin{equation}
    \label{equ:minStrokeMonitoring}
  \tilde{\sigma} = \arg \min_{\tilde{\sigma}} \left\{ \norm{\tilde{L}_e ( \tilde{V} - \tilde{U}(\tilde{\sigma}) )}^2 + p(\tilde{\sigma}) \right\},
\end{equation}
where the diagonal blocks of $\tilde{L}_e$ contain the Cholesky factors of the noise precision matrices of measurements $V_1$ and $V_2$,
\begin{align}
    \tilde{V} = \begin{bmatrix} V_1 \\ V_2 \end{bmatrix},\quad &
    \tilde{U} = \begin{bmatrix} U(\sigma_1) \\ U(\sigma_1 + K\delta\sigma) \end{bmatrix}, \\ 
    \tilde{\sigma} = \begin{bmatrix} \sigma_1\\\delta\sigma \end{bmatrix}, \quad & 
    \tilde{e} = \begin{bmatrix} e_1\\ e_2 \end{bmatrix},
\end{align}
and the regularization functional
\begin{equation}
    \label{equ:pOur}
     p(\tilde{\sigma}) = p_{\delta\sigma}(\delta\sigma) + p_{\sigma_1}(\sigma_1)
\end{equation}
allows independent regularization models for $\delta \sigma$ and $\sigma_1$.

The regularization model 
for $\delta \sigma$ is chosen to be the smoothed total variation regularizer \cite{Rudin1992}
\begin{align}
  \label{equ:TVds}
  p_{\delta\sigma}(\delta\sigma) = TV(\delta \sigma) = \alpha_{\delta\sigma} \int_{\Omega} \left( \norm{\nabla \delta \sigma}^2 + \beta^2 \right)^{1/2} {\rm d}x,
\end{align}
where $\alpha_{\delta \sigma} > 0$ is the regularization weight coefficient, $\nabla \delta \sigma$ is the gradient of the conductivity change, and $\beta > 0$ is a small smoothing parameter that ensures differentiability. This is a reasonable choice since the conductivity change caused by stroke expansion is due to a localized change of blood volume. 

The conductivity $\sigma_1$, the baseline conductivity at measurement time $t_1$, is expected to correlate well with the structure of the patient CT. This information is utilized via the parallel level sets based, spatially and directionally weighted, smoothed TV regularization that promotes similar alignment of level sets in $\sigma_1$ and the CT based reference image \cite{kolehmainen2019ipi}, giving
\begin{equation}
    \label{equ:LSWTV}
    p_{\sigma_1}(\sigma_1) = WTV(\sigma_1) = \alpha_{\sigma_1} \int_{\Omega} \left( \norm{\nabla\sigma_1}_{B(\kappa)}^2  + \beta^2\right)^{1/2} {\rm d}x,
\end{equation}
where $\alpha_{\sigma_1} > 0$ is the regularization weight coefficient, $\kappa(x)$ is the reference image, and the tensor $B(\kappa)$ is chosen such that aligned edges (or level sets) in $\sigma_1(x)$ and the reference image $\kappa(x)$ are promoted. 
The solution of \eqref{equ:minStrokeMonitoring} is based on a Lagged Gauss-Newton iteration equipped with a line search algorithm.
For more details on the implementation, see \cite{kolehmainen2019ipi, Toivanen21, toivanen2024simulation}.  Figure~\ref{fig:hammer} (right) shows a sample reconstruction produced using this MO method in 5 hours and 3 minutes.

\begin{figure}[t!]
    \centering
    \includegraphics[width=\linewidth]{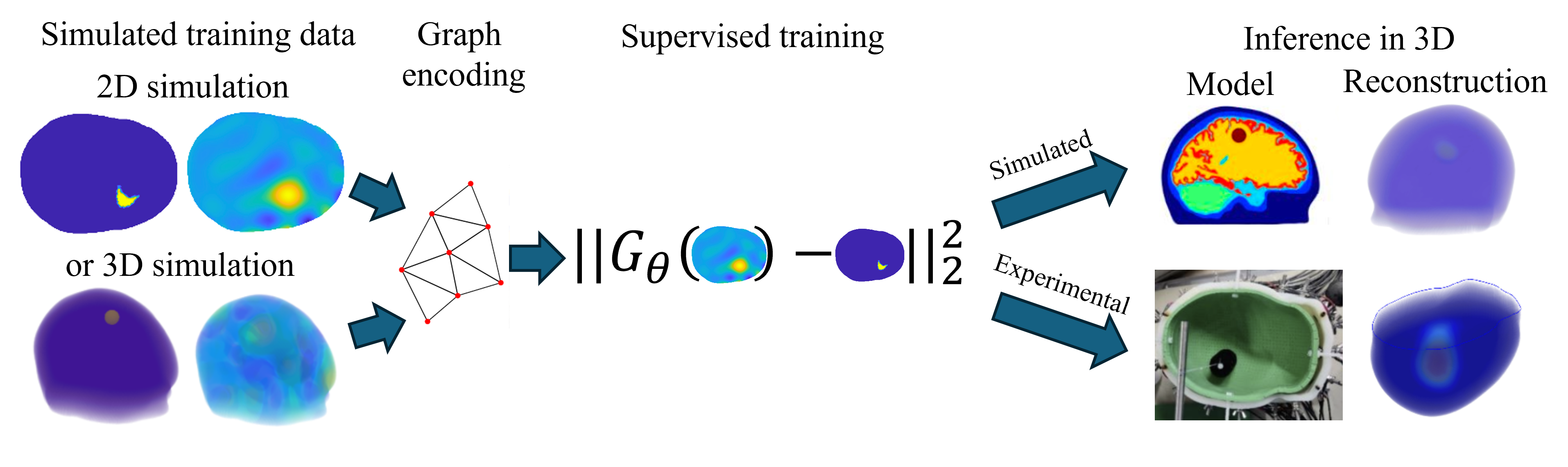}
    \caption{Graph convolutional networks can be trained on 2D or 3D input data. The graph U-net is trained on linear difference (LD) reconstructions, and then applied to linear difference images from simulated and experimental voltages. The resulting processed reconstruction is shown on the right.}
    \label{fig:overview}
\end{figure}

\subsection{Graph convolutional networks (GCNs)}
Following the work of \cite{herzberg2023domain}, we chose to work with graph convolutional neural networks instead of the typical convolutional neural networks (CNNs).  We opted for the graph setting as our reconstructed quantity, the internal conductivity change, is recovered over an irregular finite element mesh.  While we could interpolate our result to a classic pixel/voxel grid, perform the learning with CNNs, and interpolate back to the finite element mesh, the natural setting for our work is the irregular mesh.  Furthermore, performing the learning on graphs allows even greater flexibility to process reconstructions coming from different dimensions, i.e. 3D data vs 2D, a feat not naturally possible for the CNN setting where the learned filters are specific to the dimension of the data (e.g., 3x3 convolutional kernels for 2D or 3x3x3 convolutional kernels for the 3D setting).  Additionally, the number of learned parameters can be significantly lower in the graph setting requiring approximately 327 thousand parameters, whereas a classic CNN U-Net has around 30 million parameters in 2D, and 90 million in 3D. 

Following \cite{herzberg2023domain}, three kMax pooling layers, each with a 7/8 reduction in the number of nodes, were used and three graph convolutions \cite{KipfWelling2017} performed per layer with 32 channels employed at the top layer and 256 at the bottom layer.  The clone-cluster unpooling was used for the decoder side of the network and skip-connections utilized.  The output of the network is defined over the input graph, determined by the given adjacency matrix $\mathcal{A}$, with no change in dimension.  
As the (sparse) adjacency matrix and graph structure are sent into the trained network as inputs, this allows the network to be dimension-independent.  The interested reader is referred to \cite{herzberg2023domain,HerzbergThesis2022} for additional details.

\subsection{Improving reconstructions with post-processing}
In this work we are adapting the post-processing approach to learned image reconstruction. Specifically, here we are given the computationally cheap reconstruction from the linear difference imaging $\hat{\delta\sigma}$, which suffers from strong blurring and noise artifacts. We aim to recover the ground-truth difference image $\delta\sigma$ by training a post-processing network $G_\theta$ to improve the initial reconstructions. This is done by minimizing the mean squared error for optimal parameters $\theta^*$ over a set of training data $\left\{\hat{\delta\sigma}^i,\delta\sigma^i\right\}$, that is 
\begin{equation}\label{eq:loss}
    \theta^* = \arg\min_\theta \sum_i \|G_\theta(\hat{\delta\sigma}^i)-\delta\sigma^i\|_2^2.
\end{equation}
As outlined above, the network architecture is chosen as the GCN U-net from \cite{herzberg2023domain}. This allows for domain-independent training and especially for training on 2D simulations and testing on 3D experimental data  as will be described in the following section. The workflow for the training and inference is illustrated in Figure~\ref{fig:overview}. 

\section{Samples and metrics}\label{sec:sampelsMetrics}

\subsection{Learning samples}\label{sec:LearningSamples}

The 2D and 3D linear difference imaging samples for the networks were obtained as follows. First, a large number of computational head phantoms with initial and expanded hemorrhages were created and subsequently used to simulate EIT measurement datasets by solving the EIT forward problem \eqref{equ:ForwardModel1}-\eqref{equ:ForwardModel5}. The voltage datasets were then used with the linear difference (LD) imaging algorithm described in Section~\ref{sec:methods-linear}.

For each head phantom, biologically appropriate conductivity values at 1~kHz frequency from \cite{gabriel1996dielectric} were used in both 2D and 3D. The value 0.06948~S/m was used for the skin layer and brain, 0.009~S/m for the skull and 0.312~S/m for the hemorrhage. We note that the conductivity value used for the hemorrhage is lower than the conductivity of blood, 0.7~S/m, as was used also in \cite{Toivanen21}, in an effort to avoid creating an unrealistically strong signal from the hemorrhage. The initial hemorrhage was simulated as a filled circle in 2D and as a ball in 3D, both with radius randomly chosen from the range [1, 2.33]cm. The expanded hemorrhage was simulated as a combination of the initial hemorrhage and an additional half-ellipse in 2D or half-ellipsoid in 3D, both with a randomly chosen main axis direction and main axis length randomly chosen from the range [1.5-7.47]cm. The simulated hemorrhage was rejected if it overlapped the skull. To create more variability, the conductivity values used for skin, brain and skull were randomly adjusted by 0 to $\pm$ 25\%. In total, 4,000 head phantoms were created in 2D and 12,000 in 3D, corresponding to 2,000, and 6,000, initial $V_1$ and expanded hemorrhage $V_2$ voltage data pairs.  
Noisy realizations of these simulated voltage measurements were obtained by adding Gaussian zero-mean random noise with standard deviation of 0.68~mV to the simulated noiseless measurements. The standard deviation of the noise was 0.067\% of the maximum amplitude of the noiseless voltages and corresponds to the approximated relative noise level of the prototype stroke measurement device from \cite{Toivanen21}. 

Linear difference imaging estimates were computed using the simulated data by solving the minimization problem \eqref{equ:minLD}: 2000 in 2D, 6000 in 3D.  To build the regularization matrix $R$, the conductivity values were assumed to correlate within a distance corresponding to one third of the width of the head model.

The FE meshes used for data simulation and the forward model of the inverse problem were dense and had refinement near the electrodes for sufficient accuracy of the FEM solution of the model \eqref{equ:ForwardModel1}-\eqref{equ:ForwardModel5}. In the inverse problem, the conductivity was approximated in a separate, coarser mesh for suitably uniform resolution representation of the unknown image. More details on the computational meshes used in 2D and 3D are shown in Table \ref{tab:MeshInfo}. 

\subsection{Independent testing samples}\label{sec:TestingSamples}

The proposed approach, processing noisy/blurry LD images with a graph U-net, was tested with simulated data from a realistic 3D six-layer head model \cite{Paldanius21}, as well as experimental data from a laboratory experiment with a 3D-printed head tank and skull.

The simulated head model EIT measurements were computed using the electric currents interface in COMSOL following \cite{Paldanius21, toivanen2024simulation}. The different tissues in the head model were simulated using the conductivity and permittivity values shown in Table \ref{table:test_sample_conds} that originate from \cite{gabriel1996dielectric}. Four different volume spherical cortical hemorrhages with identical center points and diameters of 15, 20, 25 and 30~mm were simulated. The 32 circular electrodes with a diameter of 1~cm on the scalp were used for 32 current injections of 1~mA at 1~kHz frequency and the resulting electrode voltages were recorded. Noisy realizations of the recorded voltages were obtained by adding Gaussian zero mean random noise with standard deviation equal to $0.067\%$ of the maximum amplitude of the noiseless voltages. The noisy voltage data were paired to create four stroke monitoring test cases with volumetric changes of the hemorrhage of 0.00~ml (20 to 20~mm diameter), 2.42~ml (15 to 20~mm diameter), 3.99~ml (20 to 25~mm diameter) and 5.96~ml (25 to 30~mm diameter). Details of the computational meshes used with the head model can be found in Table \ref{tab:MeshInfo}, and more details of the head model and data simulation can be found in \cite{Paldanius21, toivanen2024simulation}.

The experimental data was measured using the KIT5 stroke EIT prototype device \cite{Toivanen21}. The measurement setup used a 3D-printed geometrically realistic adult head shaped tank, and a geometrically and electrically realistic 3D-printed skull, both based on \cite{Avery2017}. The head tank was filled with saline of conductivity 0.40~S/m and hemorrhages of different volumes were created by suspending conductive 3D printed cylinders in the saline. The cylinders all had conductivity of 4.73~S/m, height of 54~mm, and diameters of 24, 30 and 40~mm. The 32 electrodes with a diameter of 1~cm were used for 32 pairwise current injections of 1~mA at 12~kHz frequency and the resulting electrode voltages were recorded. The measured voltage data were paired to create six stroke monitoring test cases with 0.00~ml (30 to 30~mm diameter), 24.43~ml (0 to 24~mm diameter), 38.17~ml (0 to 30~mm diameter), 67.86~ml (0 to 40~mm diameter), 29.69~ml (24 to 30~mm diameter) and 13.74~ml (30 to 40~mm diameter) hemorrhage expansions. The measured conductivity values for the experimental data test case are shown in Table \ref{table:test_sample_conds} and details of the computational meshes can be found in Table \ref{tab:MeshInfo}.

\begin{table}[htbp]
\caption{Number of nodes and elements in the computational meshes used for data simulation,  computation of voltages, and discretization of the unknown conductivity for solving the inverse problem.}
\label{tab:MeshInfo}
\begin{tabular}{lrrr}
  \hline
  \hline
   & Data simulation & Voltages & Conductivity \\
  \hline
  \underline{2D training} & & &\\
  Nodes & 14,873 & 10,195 & 7,187 \\
  Elements & 29,078 & 19,823 & 14,021 \\
  \hline
  \underline{3D training} & & &\\
  Nodes & 116,547 & 67,651 & 36,496 \\
  Elements & 634,459 & 349,487 & 197,422 \\
  \hline
  \underline{Head model} & & &\\
  Nodes & $\sim$ 400k & 116,235 & 38,433 \\
  Elements & $\sim$2.5M & 597,631 & 207,453 \\
  \hline
  \underline{Experimental} & & &\\
  Nodes & n/a & 77,840 & 57,526 \\
  Elements & n/a & 417,205 & 316,788 \\
 \hline
 \hline
\end{tabular}
\end{table}

\begin{table}[htb]
\centering
\caption{Conductivity and permittivity values for all datasets.}
\footnotesize
\begin{tabular}{r||r||rr||r}
          & Training & \multicolumn{2}{c}{Head model} & Experimental \\
   Tissue & $\sigma$ (S/m) & $\sigma$ (S/m) & $\epsilon$ (F/m) & $\sigma$ (S/m)\\
   \hline
   Scalp & 0.06 & 0.32 & 434932 & 0.40 \\
   Skull & 0.009 & 0.02 & 2702 & $\sim$0.02-0.13 \\
   CSF & n/a & 2.00 & 109 & n/a \\
   White matter & 0.06 & 0.06 & 69810 & 0.40  \\
   Gray matter & 0.06 & 0.10 & 164062 & 0.40 \\
   Cerebellum & n/a & 0.12 & 164358 & n/a  \\
   Blood & 0.312 & 0.70 & 5259 & 4.73 \\
\end{tabular}
\label{table:test_sample_conds}
\end{table}

\subsection{Metrics}\label{sec:metrics}
The quality of the input reconstructions, the network processed reconstructions, and the reference reconstructions from the stroke monitoring algorithm were evaluated using the following metrics. The overall point-wise correspondence of reconstructions and truths was measured using the mean square error (MSE) and the peak signal to noise ratio (PSNR). The localization of the hemorrhage expansion was measured using the center of mass distance between reconstructions and the truths using a thresholding level corresponding to half the maximum value for each reconstruction separately. The volume errors between the reconstructions and the truths were evaluated based on volumes determined by thresholding with half the maximum value for each reconstruction separately.

\section{Results}\label{sec:results}
\begin{figure}[htbp]
\linethickness{.3mm}
    \centering
    \includegraphics[width=0.96\columnwidth]{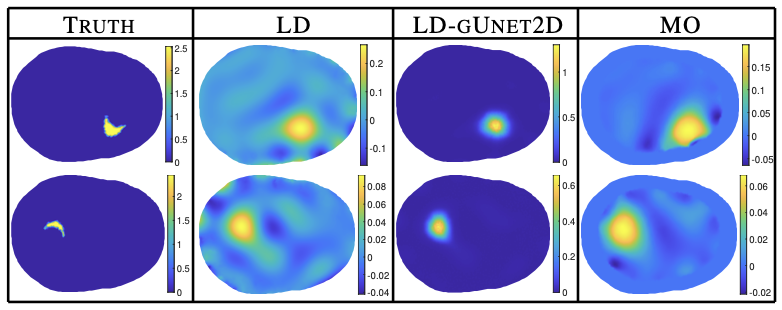}
%
%
%
%
%
%
%
\caption{Results for processing simulated LD reconstructions, consistent with training/validation data, through the LD-gUnet2D network.}
    \label{fig:2dresults-sims}
\end{figure}

\begin{figure*}[htbp]
    \centering
    \includegraphics[width=2\columnwidth]{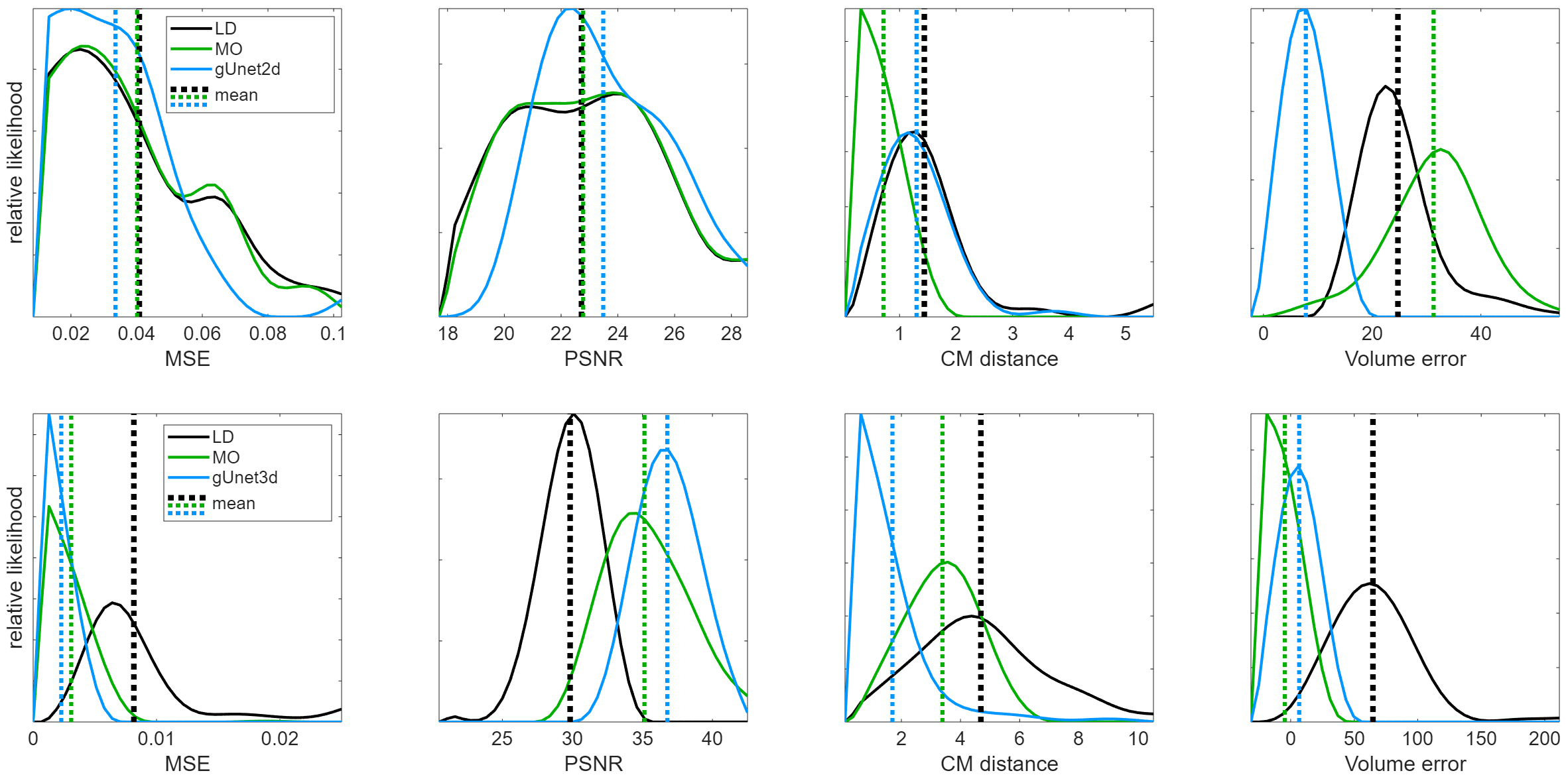}
    \caption{Estimated probability density functions and their means for the error metrics for the original linear difference imaging reconstructions (LD), the reference stroke monitoring reconstructions (MO), and for the 2D and 3D network processed linear difference imaging samples (gUnet2d, top row of images; gUnet3d, bottom row of images) using data consistent with training and validation.} 
    \label{fig:2dand3dresults-pdfs}
\end{figure*}

We will explore how well our 2D and 3D trained graph U-nets generalize to unseen 3D data, and compare them to a reliable, but computationally costly, nonlinear reconstruction method (MO), but first we verify that the networks perform well on data similar to the training data. Figure~\ref{fig:2dresults-sims} shows the result of the 2D graph U-net (gUnet2D) network on 2D simulated data consistent with, but not included in, the training data.  Unsurprisingly, the networks perform very well on data similar to the training data.  The results were analogous for the 3D network on the 3D data that was similar to training.  Figure~\ref{fig:2dand3dresults-pdfs} directly compares the estimated probability density functions of the error metrics using 200 test data samples, omitted from the training process, supporting the claim that the networks significantly improve the metrics for data similar to the training data. 

\begin{figure}[ht!]
\linethickness{.15mm}
    \centering
        \includegraphics[width=\columnwidth]{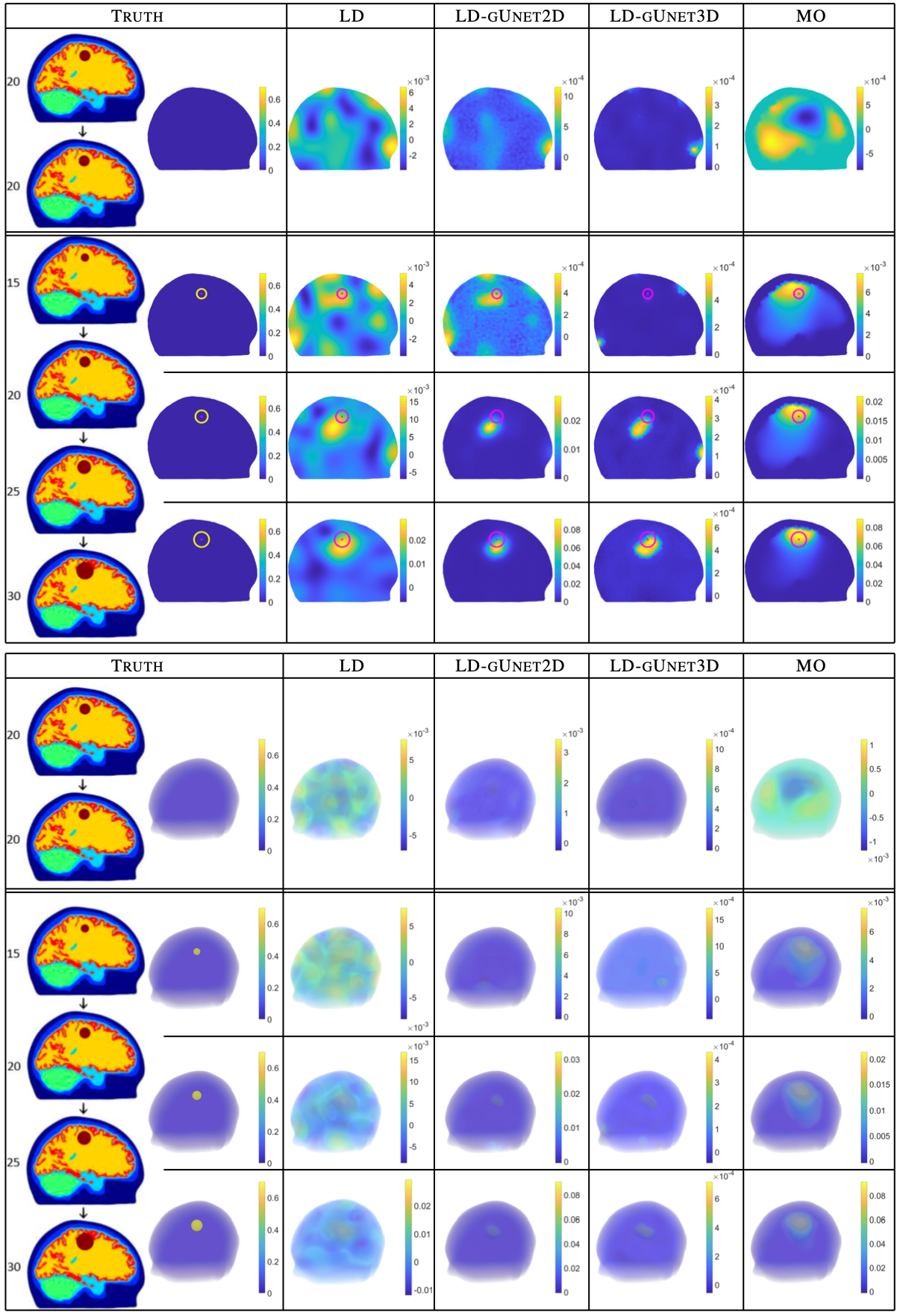}
    \caption{Computational head model test case: Results for processing the linear difference imaging reconstructions (LD) through the 2D and 3D graph-Unet networks (LD-gUnet2D and LD-gUnet3D), and the reference stroke monitoring reconstructions (MO).}    
    \label{fig:TAU-results}
\end{figure}

Next, we explore how well the networks generalize to reconstructions coming from noisy voltage data from the complex 6-layer 3D head model. Figure~\ref{fig:TAU-results} shows that the 2D and 3D graph U-nets each significantly sharpen the input (LD) images removing the many artifacts present in the LD images.  Both the 2D and 3D trained gUnets struggle with the smallest growth case (15 to 20 {\rm mm}) with the 3D network appearing to remove the growth completely; given the low-quality (LD) input image for that case, this result is not unexpected.  Directly comparing the 2D and 3D network results, we see that the 2D-trained network resulted in images with higher contrast (on par with the MO method) and fewer artifacts. The scales on all of the cases are nearly the same for the 3D network, which presents a challenge when this means the artifacts are the same magnitude as the true inclusions. The spherical shape of the recovered spherical shell change is better recovered in the 2D network as well. Error metrics for the input images (LD), 2D and 3D gUnet processed images, and reference stroke monitoring algorithm are shown in Table~\ref{tab:TAU_Metrics}. The error metrics support the findings of visual inspection. The values of the point-wise computed MSE and PSNR for the input images are strongly affected by the spherical shell shape of the true conductivity changes that is not recoverable by the linear difference imaging algorithm. However, both the 2D and the 3D networks tend to improve or keep the MSE and PSNR metrics approximately the same, except for two cases where the 3D network worsens the MSE value significantly (3.99 ml and 5.96 ml volume changes). The center of mass error metric shows that the networks do not significantly change the location of the estimated conductivity change, except in the 2.42 ml case where the input image is very poor. The artifact removal effect of both networks is obvious from volume error values which improved for all cases.  Note that the networks were only trained on directional expansions, not spherical expansions.

\begin{table}[htbp]
    \centering
    \caption{Error metrics for the computational head model out-of-distribution test case. The first column shows the true diameters of the simulated hemorrhage blood spheres ($\varnothing_1$-$\varnothing_2$). The following columns show the error metrics for the input images (LD), the 2D and 3D gUnet processed images, the reference stroke monitoring images (MO). Improvements over the LD images are highlighted in green and the best value for each case is shown in bold.}
  
 \begin{tabular}{rrrrr}
    $\varnothing_1$-$\varnothing_2$ [mm] & LD & $\Delta$gUnet2d & $\Delta$gUnet3d & $\Delta$MO \\
    \hline
    & \multicolumn{4}{l}{MSE ($\times 10^{-6}$) [S$^2$/m$^2$] \hrulefill} \\
    20-20 &    4.34 & \color{Green}   0.055 & \color{Green} \textbf{0.0013} & \color{Green}    0.074\\
    15-20 &  481.21 & \color{Green} 479.53  & \color{Green} 479.57   & \color{Green}  \textbf{472.59}\\
    20-25 &  841.50 &               844.68  &               852.40   & \color{Green}  \textbf{819.64}\\
    25-30 & 1124.00 & \color{Green}1097.78  &              1171.30   & \color{Green} \textbf{1039.60}\\
    & \multicolumn{4}{l}{PSNR [dB] \hrulefill} \\
    20-20 & n/a   & n/a   & n/a   & n/a\\
    15-20 & 29.30 & \color{Green} 29.32 & \color{Green} 29.32 & \color{Green} \textbf{29.38}\\
    20-25 & 26.87 &               26.86 &               26.82 & \color{Green} \textbf{26.99}\\
    25-30 & 25.62 & \color{Green} 25.72 &               25.44 & \color{Green} \textbf{25.96}\\
    & \multicolumn{4}{l}{Center of mass error [cm] \hrulefill} \\
    20-20 & n/a  & n/a  & n/a  & n/a\\
    15-20 & 5.19 &               9.73 & 7.40 & \color{Green} \textbf{1.37}\\
    20-25 & 3.03 & \color{Green} 2.26 & 3.07 & \color{Green} \textbf{0.87}\\
    25-30 & 1.43 &               1.49 & 1.49 & \color{Green} \textbf{0.72}\\
    & \multicolumn{4}{l}{Volume error [cm$^3$] \hrulefill} \\
    20-20 & 107.59 & \color{Green}  8.04 & \color{Green}  \textbf{1.87} & \color{Green} 101.47\\
    15-20 & 238.03 & \color{Green} 10.44 & \color{Green}  \textbf{1.05} & \color{Green} 97.045\\
    20-25 &  52.93 & \color{Green}  \textbf{1.97} & \color{Green} 17.02 &               75.77\\
    25-30 &  36.09 & \color{Green} \textbf{-0.34} & \color{Green} 10.89 & \color{Green} 35.85\\
    \end{tabular}    
    \label{tab:TAU_Metrics}
\end{table}

%

\begin{figure}
\linethickness{.15mm}
    \centering
            \includegraphics[width=\columnwidth]{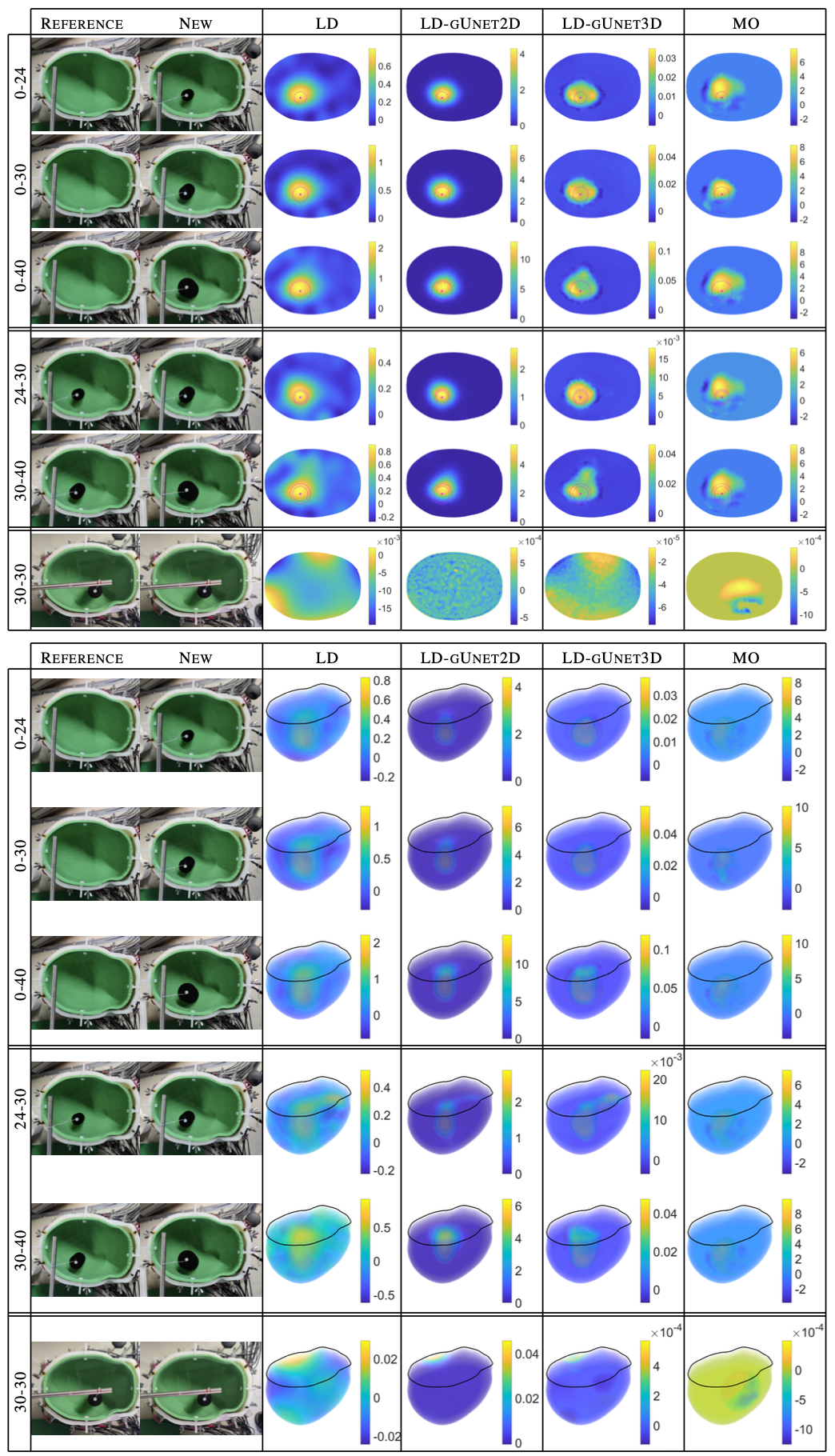}
    \caption{Laboratory test case: Results for processing the linear difference imaging reconstructions (LD) through the 2D and 3D graph-Unet networks (LD-gUnet2D and LD-gUnet3D), and the reference stroke monitoring reconstructions (MO).}
    \label{fig:LAB-results}
\end{figure}

Lastly, we explore how well the networks generalize to the experimental lab data.  Figure~\ref{fig:LAB-results} compares the results of the 2D and 3D gUnets to those from the nonlinear stroke monitoring algorithm (MO) for the 3D reconstructions from the experimental lab data.  Both networks produce significantly sharper images with less prominent artifacts than both the input linear reconstructions (left) and nonlinear, computationally costly, stroke monitoring algorithm images (right).  This is particularly pronounced when looking at the 3D volume plots.  The targets are easily visible in the post-processed gUnet columns whereas there are common artifacts surrounding the targets in the far right column.  


\section{Discussion}\label{sec:discussion}

The results suggest that the proposed gUnet post-processing of the LD reconstruction leads to a similar performance in hemorrhage growth detection as the non-linear stroke monitoring algorithm (MO), but in a tiny fraction of computational time (a few minutes compared to several hours), making it a potential candidate to a clinical on-line monitoring setup where fast reconstruction is needed. 

\begin{figure*}[t]
    \centering
                \includegraphics[width=\textwidth]{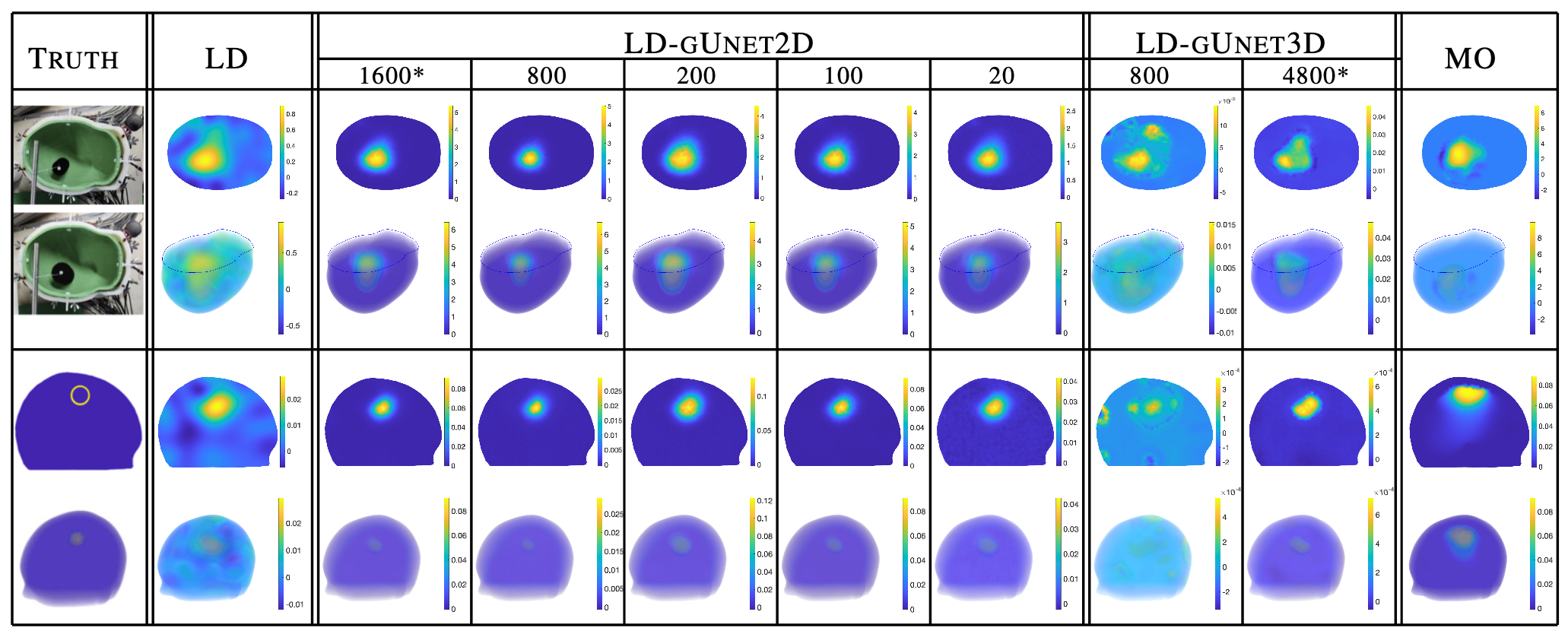}
    \caption{Comparison of performance of the graph U-nets trained using 2D and 3D samples when fewer training samples are used.  Columns with an asterisk~* represent the networks used in the rest of the paper.  The top two rows show the experimental lab dataset simulating a medium bleed (30mm diameter) increasing to a larger (40mm diameter) bleed.  The bottom two rows show results for the simulated six-layer head model with the change between a 25 and 30mm bleed.  Slice views (top) and volume views (bottom) are shown.  Each figure is on its own scale.}
    \label{fig:varying_levels_training}
\end{figure*}

A surprising result in this project was the ability to train the 2D gUnets on very few samples.  Figure~\ref{fig:varying_levels_training} compares results for the 2D and 3D gUnet networks with varying numbers of training data. With as few as 20 training samples, and 4 validation samples, the 2D gUnet produced remarkable results with only a slight degradation in image quality and recovered magnitudes. By contrast, even 800 training, and 180 validation samples in 3D proved inadequate. 
	
\begin{figure*}[t]
    \centering
                    \includegraphics[width=\textwidth]{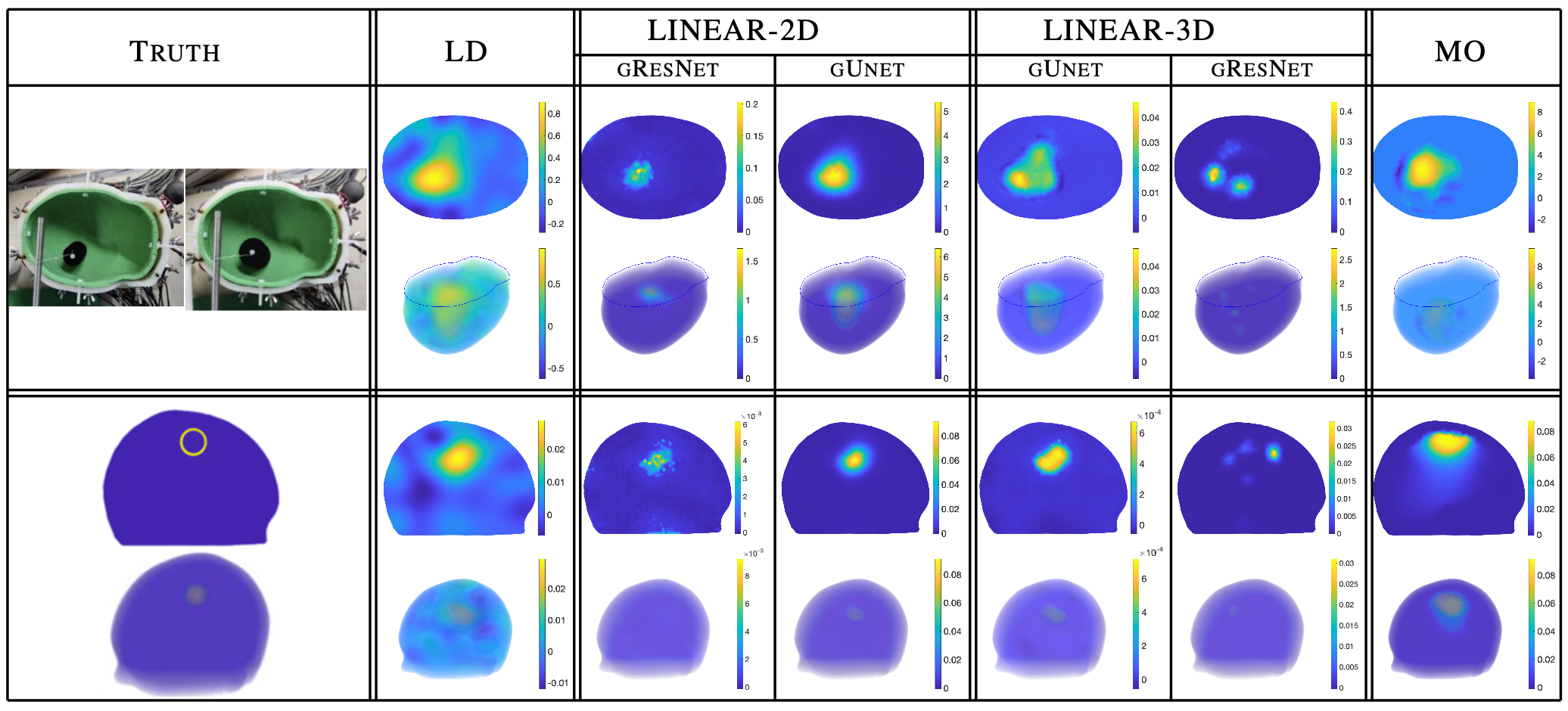}
    \caption{Comparison of the performance of the Graph Residual Networks ({\sc gResNet}) to the Graph U-nets ({\sc gUnet}), in 2D and 3D.  The top two rows again show the experimental lab dataset where a medium bleed grows to a larger bleed (slice, volume). The bottom two rows show the results for a 25mm bleed growing to a 30mm bleed on the simulated six-layer head model (slice, volume).  Each figure is on its own scale.}
    \label{fig:gUnet_vs_Resnet}
\end{figure*}

A natural question is whether a U-net was necessary for this task.  Could a simpler ResNet suffice?  Following the approach of \cite{herzberg2021graph}, we compared to a Graph Residual Network (denoted here as gResNet), with 10~residual blocks and 3~graph convolutions per block. While there are alternative choices for the gResNet structure, and tweaks on the numbers of parameters that could be performed,  in our testing, the gResNets performed adequately on data similar to the training data, but struggled to generalize to the more complicated six-layer head model and experimental data.  Figure~\ref{fig:gUnet_vs_Resnet} shows the speckling/pixelated appearance in the images, whereas the gUnet images appear more clearly defined. 

\subsection{Computational Cost}
\subsubsection{Creating training data}
The average combined computation time for the creation of a single head phantom pair and the corresponding data simulation with both head phantoms was 0.30~s in 2D and 26.60~s in 3D. Obtaining the linear difference imaging reconstructions, by solving the minimization problem \eqref{equ:minLD}, took on average 0.93~s in 2D and 35.42~s in 3D. In both cases, the regularization matrix in \eqref{equ:minLD} was precomputed once and re-used with all measurement sets. This precomputation time is not included in the previously stated times, and was 3.30~s in 2D and 87.20~s in 3D. The training data was created on CSC Finland's ePouta cloud server partition with 232 GiB of memory and 64 AMD EPYC processors.

\subsubsection{Network training times}
The networks were trained using the supervised learning approach \eqref{eq:loss}, with 1600 samples used for training in 2D, 4800 in 3D, and 300 for validation in 2D and 1200 for validation in 3D.  Each network was trained using a patience of 50 epochs and the Adam optimizer with an initial learning rate of 0.001.  Network training times, approximately, were as follows: 2D~gUnet~(6~sec/epoch, 17~min) and 2D~gResNet~(17~sec/epoch, 105~min), 3D~gUnet~(2~min/epoch, 380~min) and 3D~gResNet~(7.5~min/epoch, 2527.5~min). Multiple instances of the training were performed with negligible changes in training times and results. The learning was performed in PyTorch version {\tt 2.2.0+cu121} and timings correspond to a workstation with an A6000 NVIDIA card.

\subsubsection{Test data reconstruction times}
For the laboratory data test case, a single linear difference (LD) imaging reconstruction was computed in approximately 3~min and a single reference stroke monitoring (MO) algorithm reconstruction in 2.5~h. For the computational head model test case, a single linear difference (LD) imaging reconstruction was computed in approximately 50~s, and a single reference stroke monitoring algorithm reconstruction in 5~h. Note that post-processing a reconstruction through the trained graph network adds a fraction of a second.



\section{Conclusions}\label{sec:conclusion}
In this study, we demonstrate the ability to use graph U-nets to post-process relatively fast, standard, linear difference (LD) reconstructions for EIT to produce improved images of comparable quality to a time-intensive nonlinear stroke-monitoring (MO) reconstruction method, at a tremendous savings in computational cost.  By training in the graph setting, instead of the classic CNN framework, we were able to perform the training on 2D images and then test on 3D data.  This was advantageous for our imaging problem, as generating the 2D training data was nearly 50x 
faster, per sample, than the corresponding 3D training data without even accounting for the higher number of training samples needed in 3D. The graph U-nets were remarkably robust to out-of-distribution training data, generalizing very well to stroke monitoring data from a complicated six-layer head model, as well as experimental lab data.

These findings suggest that the proposed reconstruction pipeline of using linear difference imaging, followed by graph U-net post processing, could be a feasible reconstruction approach for practical on-line monitoring where fast reconstructions are needed. 

\section*{Acknowledgment}\label{sec:acknowledgments}
The authors wish to acknowledge CSC - IT Center for Science, Finland, for computational resources.



\bibliographystyle{IEEEtran}
\small
\bibliography{bibrefs}

\begin{thebibliography}{10}
\providecommand{\url}[1]{#1}
\csname url@samestyle\endcsname
\providecommand{\newblock}{\relax}
\providecommand{\bibinfo}[2]{#2}
\providecommand{\BIBentrySTDinterwordspacing}{\spaceskip=0pt\relax}
\providecommand{\BIBentryALTinterwordstretchfactor}{4}
\providecommand{\BIBentryALTinterwordspacing}{\spaceskip=\fontdimen2\font plus
\BIBentryALTinterwordstretchfactor\fontdimen3\font minus
  \fontdimen4\font\relax}
\providecommand{\BIBforeignlanguage}[2]{{%
\expandafter\ifx\csname l@#1\endcsname\relax
\typeout{** WARNING: IEEEtran.bst: No hyphenation pattern has been}%
\typeout{** loaded for the language `#1'. Using the pattern for}%
\typeout{** the default language instead.}%
\else
\language=\csname l@#1\endcsname
\fi
#2}}
\providecommand{\BIBdecl}{\relax}
\BIBdecl

\bibitem{katan2018global}
M.~Katan and A.~Luft, ``Global burden of stroke,'' in \emph{Seminars in
  neurology}, vol.~38, no.~02.\hskip 1em plus 0.5em minus 0.4em\relax Thieme
  Medical Publishers, 2018, pp. 208--211, [doi: 10.1055/s-0038-1649503].

\bibitem{greenberg20222}
S.~M. Greenberg, W.~C. Ziai, C.~Cordonnier, D.~Dowlatshahi, B.~Francis, J.~N.
  Goldstein, J.~C. Hemphill~III, R.~Johnson, K.~M. Keigher, W.~J. Mack
  \emph{et~al.}, ``2022 guideline for the management of patients with
  spontaneous intracerebral hemorrhage: a guideline from the american heart
  association/american stroke association,'' \emph{Stroke}, vol.~53, no.~7, pp.
  e282--e361, 2022, [doi: 10.1161/STR.0000000000000407].

\bibitem{sussman2013hemorrhagic}
E.~S. Sussman and E.~S. Connolly~Jr, ``Hemorrhagic transformation: a review of
  the rate of hemorrhage in the major clinical trials of acute ischemic
  stroke,'' \emph{Frontiers in neurology}, vol.~4, p.~69, 2013.

\bibitem{malone2014}
\BIBentryALTinterwordspacing
E.~Malone, M.~Jehl, S.~Arridge, T.~Betcke, and D.~Holder, ``Stroke type
  differentiation using spectrally constrained multifrequency {EIT}: evaluation
  of feasibility in a realistic head model,'' \emph{Physiological Measurement},
  vol.~35, no.~6, pp. 1051--1066, may 2014, [doi: 10.1088/0967-3334/35/6/1051].
  [Online]. Available:
  \url{https://doi.org/10.1088\%2F0967-3334\%2F35\%2F6\%2F1051}
\BIBentrySTDinterwordspacing

\bibitem{goren2018}
N.~Goren, J.~Avery, T.~Dowrick, E.~Mackle, A.~Witkowska-Wrobel, D.~Werring, and
  D.~Holder, ``Multi-frequency electrical impedance tomography and neuroimaging
  data in stroke patients,'' \emph{Scientific data}, vol.~5, no.~1, pp. 1--10,
  2018, [doi: 10.1038/sdata.2018.112].

\bibitem{horesh2006thesis}
L.~Horesh, \emph{Some novel approaches in modelling and image reconstruction
  for multi frequency electrical impedance tomography of the human
  brain}.\hskip 1em plus 0.5em minus 0.4em\relax University of London, 2006.

\bibitem{candiani2021}
\BIBentryALTinterwordspacing
V.~Candiani, N.~Hyv\"onen, J.~P. Kaipio, and V.~Kolehmainen, ``Approximation
  error method for imaging the human head by electrical impedance tomography,''
  \emph{Inverse Problems}, vol.~37, no.~12, p. 125008, 2021, [doi:
  10.1088/1361-6420/ac346a]. [Online]. Available:
  \url{https://doi.org/10.1088/1361-6420/ac346a}
\BIBentrySTDinterwordspacing

\bibitem{agnelli2020classification}
J.~P. Agnelli, A.~{\c{C}}{\"o}l, M.~Lassas, R.~Murthy, M.~Santacesaria, and
  S.~Siltanen, ``Classification of stroke using neural networks in electrical
  impedance tomography,'' \emph{Inverse Problems}, vol.~36, no.~11, p. 115008,
  2020, [doi: 10.1088/1361-6420/abbdcd].

\bibitem{mcdermott2020}
B.~McDermott, A.~Elahi, A.~Santorelli, M.~O’Halloran, J.~Avery, and
  E.~Porter, ``Multi-frequency symmetry difference electrical impedance
  tomography with machine learning for human stroke diagnosis,''
  \emph{Physiological Measurement}, vol.~41, no.~7, p. 075010, 2020, [doi:
  10.1088/1361-6579/ab9e54].

\bibitem{candiani2022}
V.~Candiani and M.~Santacesaria, ``Neural networks for classification of stroke
  in electrical impedance tomography on a 3{D} head model,'' \emph{Mathematics
  in Engineering}, vol.~4, no.~4, pp. 1--22, 2022, [doi: 10.3934/mine.2022029].

\bibitem{shi2021sparse}
Y.~Shi, Y.~Wu, M.~Wang, Z.~Tian, X.~Kong, and X.~He, ``Sparse image
  reconstruction of intracerebral hemorrhage with electrical impedance
  tomography,'' \emph{Journal of Medical Imaging}, vol.~8, no.~1, pp.
  014\,501--014\,501, 2021, [doi: 10.1117/1.JMI.8.1.014501].

\bibitem{xu2010}
\BIBentryALTinterwordspacing
C.-H. Xu, L.~Wang, X.-T. Shi, F.-S. You, F.~Fu, R.-G. Liu, M.~Dai, Z.-W. Zhao,
  G.-D. Gao, and X.-Z. Dong, ``Real-time imaging and detection of intracranial
  haemorrhage by electrical impedance tomography in a piglet model,''
  \emph{Journal of International Medical Research}, vol.~38, no.~5, pp.
  1596--1604, 2010, [doi: 10.1177/147323001003800504]. [Online]. Available:
  \url{https://doi.org/10.1177/147323001003800504}
\BIBentrySTDinterwordspacing

\bibitem{dai2013}
M.~Dai, B.~Li, S.~Hu, C.~Xu, B.~Yang, J.~Li, F.~Fu, Z.~Fei, and X.~Dong, ``In
  vivo imaging of twist drill drainage for subdural hematoma: a clinical
  feasibility study on electrical impedance tomography for measuring
  intracranial bleeding in humans,'' \emph{PloS one}, vol.~8, no.~1, p. e55020,
  2013, [doi: 10.1371/journal.pone.0055020].

\bibitem{yang2019}
B.~Yang, B.~Li, C.~Xu, S.~Hu, M.~Dai, J.~Xia, P.~Luo, X.~Shi, Z.~Zhao, X.~Dong
  \emph{et~al.}, ``Comparison of electrical impedance tomography and
  intracranial pressure during dehydration treatment of cerebral edema,''
  \emph{NeuroImage: Clinical}, vol.~23, p. 101909, 2019, [doi:
  10.1016/j.nicl.2019.101909 ].

\bibitem{toivanen2024simulation}
J.~Toivanen, A.~Paldanius, B.~Dekdouk, V.~Candiani, A.~H{\"a}nninen,
  T.~Savolainen, D.~Strbian, N.~Forss, N.~Hyv{\"o}nen, J.~Hyttinen
  \emph{et~al.}, ``Simulation-based feasibility study of monitoring of
  intracerebral hemorrhages and detection of secondary hemorrhages using
  electrical impedance tomography,'' \emph{Journal of Medical Imaging},
  vol.~11, no.~1, pp. 014\,502--014\,502, 2024.

\bibitem{Toivanen21}
J.~Toivanen, A.~H{\"a}nninen, T.~Savolainen, N.~Forss, and V.~Kolehmainen,
  ``Monitoring hemorrhagic strokes using {EIT},'' in \emph{Bioimpedance and
  Spectroscopy}.\hskip 1em plus 0.5em minus 0.4em\relax Elsevier, 2021, pp.
  271--298, [doi: 10.1016/B978-0-12-818614-5.00007-2].

\bibitem{tanyu2023electrical}
D.~N. Tanyu, J.~Ning, A.~Hauptmann, B.~Jin, and P.~Maass, ``Electrical
  impedance tomography: A fair comparative study on deep learning and
  analytic-based approaches,'' \emph{arXiv preprint arXiv:2310.18636}, 2023.

\bibitem{li2017image}
X.~Li, Y.~Lu, J.~Wang, X.~Dang, Q.~Wang, X.~Duan, and Y.~Sun, ``An image
  reconstruction framework based on deep neural network for electrical
  impedance tomography,'' in \emph{2017 IEEE International Conference on Image
  Processing (ICIP)}.\hskip 1em plus 0.5em minus 0.4em\relax IEEE, 2017, pp.
  3585--3589.

\bibitem{chen2022mmv}
Z.~Chen, J.~Xiang, P.-O. Bagnaninchi, and Y.~Yang, ``Mmv-net: A multiple
  measurement vector network for multifrequency electrical impedance
  tomography,'' \emph{IEEE Transactions on Neural Networks and Learning
  Systems}, vol.~34, no.~11, pp. 8938--8949, 2022.

\bibitem{Hamilton2018_DeepDbar}
S.~J. Hamilton and A.~Hauptmann, ``Deep d-bar: Real time electrical impedance
  tomography imaging with deep neural networks,'' \emph{IEEE Transactions on
  Medical Imaging}, vol.~37, no.~10, pp. 2367--2377, 2018.

\bibitem{wei2019dominant}
Z.~Wei, D.~Liu, and X.~Chen, ``Dominant-current deep learning scheme for
  electrical impedance tomography,'' \emph{IEEE Transactions on Biomedical
  Engineering}, vol.~66, no.~9, pp. 2546--2555, 2019.

\bibitem{hamilton2019beltrami}
S.~J. Hamilton, A.~H{\"a}nninen, A.~Hauptmann, and V.~Kolehmainen,
  ``Beltrami-net: domain-independent deep d-bar learning for absolute imaging
  with electrical impedance tomography (a-eit),'' \emph{Physiological
  measurement}, vol.~40, no.~7, p. 074002, 2019.

\bibitem{chen2021structure}
Z.~Chen and Y.~Yang, ``Structure-aware dual-branch network for electrical
  impedance tomography in cell culture imaging,'' \emph{IEEE Transactions on
  Instrumentation and Measurement}, vol.~70, pp. 1--9, 2021.

\bibitem{herzberg2021graph}
W.~Herzberg, D.~B. Rowe, A.~Hauptmann, and S.~J. Hamilton, ``Graph
  convolutional networks for model-based learning in nonlinear inverse
  problems,'' \emph{IEEE transactions on computational imaging}, vol.~7, pp.
  1341--1353, 2021.

\bibitem{herzberg2023domain}
W.~Herzberg, A.~Hauptmann, and S.~J. Hamilton, ``Domain independent
  post-processing with graph u-nets: applications to electrical impedance
  tomographic imaging,'' \emph{Physiological Measurement}, vol.~44, no.~12, p.
  125008, 2023.

\bibitem{Dong2024Graph}
Z.~Liu, J.~Wang, Q.~Shan, and D.~Liu, ``Grapheit: Unsupervised graph neural
  networks for electrical impedance tomography,'' \emph{IEEE Transactions on
  Computational Imaging}, pp. 1--12, 2024.

\bibitem{Paldanius21}
A.~Paldanius, B.~Dekdouk, J.~Toivanen, V.~Kolehmainen, and J.~Hyttinen,
  ``Sensitivity analysis highlights the importance of accurate head models for
  electrical impedance tomography monitoring of intracerebral hemorrhagic
  stroke,'' \emph{IEEE Transactions on Biomedical Engineering}, vol.~69, no.~4,
  pp. 1491--1501, 2021, [doi: 10.1109/TBME.2021.3120929].

\bibitem{Cheng1989}
K.~Cheng, D.~Isaacson, J.~Newell, and D.~Gisser, ``Electrode models for
  electric current computed tomography,'' \emph{IEEE Transactions on Biomedical
  Engineering}, vol.~36, no.~9, pp. 918--924, sep 1989, [doi:
  10.1109/10.35300].

\bibitem{Somersalo1992}
E.~Somersalo, M.~Cheney, and D.~Isaacson, ``Existence and uniqueness for
  electrode models for electric current computed tomography,'' \emph{SIAM
  Journal on Applied Mathematics}, vol.~52, no.~4, pp. 1023--1040, 1992, [doi:
  10.1137/0152060].

\bibitem{vauhkonen98}
M.~Vauhkonen, D.~Vad{\'a}sz, P.~A. Karjalainen, E.~Somersalo, and J.~P. Kaipio,
  ``Tikhonov regularization and prior information in electrical impedance
  tomography,'' \emph{IEEE transactions on medical imaging}, vol.~17, no.~2,
  pp. 285--293, 1998, [doi: 10.1109/42.700740].

\bibitem{Kaipio2000stat}
J.~Kaipio, V.~Kolehmainen, E.~Somersalo, and M.~Vauhkonen, ``Statistical
  inversion and {{M}onte} {{C}arlo} sampling methods in electrical impedance
  tomography,'' \emph{Inverse Problems}, vol.~16, no.~5, pp. 1487--1522, 2000,
  [doi: 10.1088/0266-5611/16/5/321].

\bibitem{barber1987fast}
D.~C. Barber and A.~D. Seagar, ``Fast reconstruction of resistance images,''
  \emph{Clinical Physics and Physiological Measurement}, vol.~8, no.~4A, p.~47,
  1987, [doi: 10.1088/0143-0815/8/4A/006].

\bibitem{Bagshaw2003}
\BIBentryALTinterwordspacing
A.~P. Bagshaw, A.~D. Liston, R.~H. Bayford, A.~Tizzard, A.~P. Gibson,
  A.~Tidswell, M.~K. Sparkes, H.~Dehghani, C.~D. Binnie, and D.~S. Holder,
  ``Electrical impedance tomography of human brain function using
  reconstruction algorithms based on the finite element method,''
  \emph{NeuroImage}, vol.~20, no.~2, pp. 752 -- 764, 2003, [doi:
  10.1016/S1053-8119(03)00301-X]. [Online]. Available:
  \url{http://www.sciencedirect.com/science/article/pii/S105381190300301X}
\BIBentrySTDinterwordspacing

\bibitem{lieberman2010parameter}
C.~Lieberman, K.~Willcox, and O.~Ghattas, ``Parameter and state model reduction
  for large-scale statistical inverse problems,'' \emph{SIAM Journal on
  Scientific Computing}, vol.~32, no.~5, pp. 2523--2542, 2010, [doi:
  10.1137/090775622].

\bibitem{Liu2015ipi}
\BIBentryALTinterwordspacing
D.~Liu, V.~Kolehmainen, S.~Siltanen, A.-M. Laukkanen, and A.~Sepp\"anen,
  ``Estimation of conductivity changes in a region of interest with electrical
  impedance tomography,'' \emph{Inverse Problems and Imaging}, vol.~9, no.~1,
  pp. 211--229, 2015, [doi: 10.3934/ipi.2015.9.211]. [Online]. Available:
  \url{/article/id/67ed07c5-f6f9-4d18-a802-0e2ffef23194}
\BIBentrySTDinterwordspacing

\bibitem{kolehmainen2019ipi}
V.~Kolehmainen, M.~J. Ehrhardt, and S.~R. Arridge, ``Incorporating structural
  prior information and sparsity into {EIT} using parallel level sets,''
  \emph{Inverse Problems \& Imaging}, vol.~13, no.~2, pp. 285--307, 2019, [doi:
  10.3934/ipi.2019015].

\bibitem{Arridge14LSQR}
S.~Arridge, M.~Betcke, and L.~Harhanen, ``Iterated preconditioned {LSQR} method
  for inverse problems on unstructured grids,'' \emph{Inverse Problems},
  vol.~30, no.~7, p. 075009, 2014, [doi: 10.1088/0266-5611/30/7/075009].

\bibitem{Harhanen15LSQR}
L.~Harhanen, N.~Hyv\"onen, H.~Majander, and S.~Staboulis, ``Edge-enhancing
  reconstruction algorithm for three-dimensional electrical impedance
  tomography,'' \emph{SIAM Journal on Scientific Computing}, vol.~37, no.~1,
  pp. B60--B78, 2015, [doi: 10.1137/140971750].

\bibitem{Rudin1992}
L.~Rudin, S.~Osher, and E.~Fatemi, ``Nonlinear total variation based noise
  removal algorithms,'' \emph{Physica D: Nonlinear Phenomena}, vol.~60, no.
  1-4, pp. 259--268, 1992, [doi: 10.1016/0167-2789(92)90242-F].

\bibitem{KipfWelling2017}
\BIBentryALTinterwordspacing
T.~N. Kipf and M.~Welling, ``Semi-supervised classification with graph
  convolutional networks,'' \emph{CoRR}, vol. abs/1609.02907, 2016. [Online].
  Available: \url{http://arxiv.org/abs/1609.02907}
\BIBentrySTDinterwordspacing

\bibitem{HerzbergThesis2022}
W.~Herzberg, ``Graph neural networks for inverse problems with flexible
  meshes,'' Ph.D. dissertation, Marquette University, 2022.

\bibitem{gabriel1996dielectric}
S.~Gabriel, R.~Lau, and C.~Gabriel, ``The dielectric properties of biological
  tissues: {III}. parametric models for the dielectric spectrum of tissues,''
  \emph{Physics in medicine \& biology}, vol.~41, no.~11, p. 2271, 1996, [doi:
  10.1088/0031-9155/41/11/003].

\bibitem{Avery2017}
\BIBentryALTinterwordspacing
J.~Avery, K.~Aristovich, B.~Low, and D.~Holder, ``Reproducible 3d printed head
  tanks for electrical impedance tomography with realistic shape and
  conductivity distribution,'' \emph{Physiological Measurement}, vol.~38,
  no.~6, pp. 1116--1131, may 2017. [Online]. Available:
  \url{https://doi.org/10.1088%2F1361-6579%2Faa6586}
\BIBentrySTDinterwordspacing

\end{thebibliography}

\end{document}